# A Physics Based Approach for Neural Networks Enabled Design of All-Dielectric Metasurfaces

Ibrahim Tanriover, Wisnu Hadibrata and Koray Aydin[*]

Department of Electrical and Computer Engineering, Northwestern University, Evanston, Illinois 60208, United States

**Abstract**
Machine learning methods have found novel application areas in various disciplines as they offer low-computational cost solutions to complex problems. Recently, metasurface design has joined among these applications, and neural networks enabled significant improvements within a short period of time. However, there are still outstanding challenges that needs to be overcome. Here, we propose a data pre-processing approach based on the governing laws of the physical problem to eliminate dimensional mismatch between high dimensional optical response and low dimensional feature space of metasurfaces. We train forward and inverse models to predict optical responses of cylindrical meta-atoms and to retrieve their geometric parameters for a desired optical response, respectively. Our approach provides accurate prediction capability even outside the training spectral range. Finally, using our inverse model, we design and demonstrate a focusing metalens as a proof-of-concept application, thus validating the capability of our proposed approach. We believe our method will pave the way towards practical learning-based models to solve more complicated photonic design problems.

Metasufaces are flat optical components that employ spatially arranged micro/nano structures as optical antennas[1], scatterers[2] or waveguides[3]. Their capability of arbitrary light manipulation within sub-wavelength dimensions, made them one of the most promising candidates to replace conventional optics. Based on the metasurface concept, various optical components, such as polarizers[4], lenses[1-3,5-12,13] and holograms[14] have been realized. Current metasurface design process can be divided into two main approaches as forward and inverse problem. The former is based on numerous trial and error sessions to obtain the desired geometrical and material parameter set, such as height, width, and depth of a rectangular scatterer or the refractive index of material to be employed, that provides the best approximation to desired optical performance. The optical response of a given parameter set is generally obtained from full electromagnetic simulations using Finite Elements Method (FEM) and Finite Difference Time Domain (FDTD) methods. Then, the designer alters these parameters with respect to the simulation results and perform new simulations for updated set or structure. This process continues until reaching desired performance criteria. Such simulations are often computationally expensive and the computational cost drastically increases with increasing structure complexity. Also, the number of required trial and error sessions is indeterminate and heavily dependent on the designer's experience. The second approach, also known as the inverse design, requires solving a computationally expensive high dimensional optimization problem, which aims to find required geometrical parameters from a given spectral response. The second approach, too, is heavily dependent on full electromagnetic simulations. These large time and computational power costs of the conventional approaches have become a limiting factor for metasurface design. This limitation is not specific only to metasurface design. Computational cost is a common problem for many areas utilizing numerical optimization. To overcome this problem, machine learning based approaches have been employed in various disciplines including chemistry[15], material science[16] and physics[17]. Recently, metasurface design also has joined among the application areas of machine learning based/assisted approaches[18-32].

Within a few years, significant progress has been made in machine learning based metasurface design approaches. The pioneer examples that are solving forward and/or inverse problem for transmission/reflection spectra of alternating dielectric layers[19,21,22] and plasmonic structures[23] are followed by solutions for 3D all-dielectric metasurfaces[20,24]. The latest examples also include solution of phase response[24-26] as well as transmission spectra. Within these pioneering works, many applications, including chirality[18], have been reported as proofs of concept. These promising results indicate a rapid progress and numerous possibilities for the future of learning-based approaches. However, there are still outstanding challenges that need to be addressed to make machine learning approaches viable and widely deployable in nanophotonic design.

One of the challenges in the forward problem of the metasurface design is the dimensional mismatch between input and output vectors of Neural Networks (NNs). A metasurface can be described by a low-dimensional feature space while the spectral response should possess sufficient number of data points to fully resolve the output features, especially the resonances. In previous studies, up and down sampling methods are used to overcome this problem[19,20-24]. However, these methods suffer from limited spectral resolution. Sacrifice of some data points is inevitable in the case of sampling, and this results in losing some of the critical portions of the data such as resonances.

Spectral generalizability is another significant challenge that needs to be addressed. Solution capability of present examples are restricted to their training spectral range. In order to solve forward (or inverse) problem for a similar parameter set (or optical response) for another spectral range, a new model should be trained for this spectral window. This requires a new training dataset to be prepared for the new spectral range, which in turn, requires significant amount of time and computational power.

In this work, utilizing the spectral scalability of dielectric metasurfaces, we propose an easy-to-implement and robust approach to solve the dimensional mismatch problem of NNs for any arbitrary spectral resolution, which also intrinsically suggests an approximated solution to spectral generalizability problem. Based on this approach, we solved both the forward and the inverse problem for transmission and phase responses of high index ($2 \leq n \leq 4$) nano cylinders on top of a low index ($n \sim 1.45$) substrate as shown in the Figure 1a. We investigate the performance of our forward model both outside and inside the spectral range of our training dataset and show that it can be used as a powerful tool even outside its training spectral range. We also construct an inverse network to provide a unit cell library that will generate the desired phase distribution, based on the proposed approach. However, an inverse problem may have many solutions resulting in an additional challenge to overcome. To solve this non-uniqueness problem, we use a modified version of previously suggested tandem learning idea[19]. Using this network, we realize on-demand design of a metalens as a proof-of-concept demonstration.

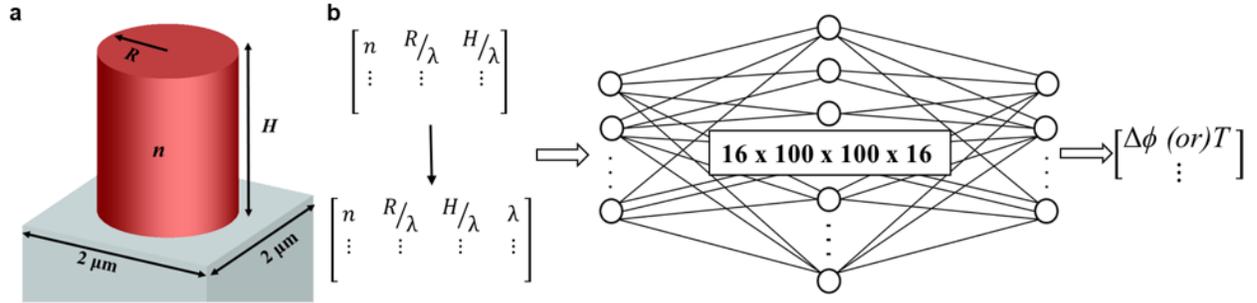

**Figure 1.** (a) The unit cell structure for investigation: cylindrical dielectric meta-atom with varying radius, height and refractive index on top of a low index ($n_{subs} \approx 1.45$) substrate. The periodicity is 2 μm, $2 \leq n \leq 4$, $300 \leq R \leq 700$ nm, and $0.75 \leq H \leq 2.50$ μm. (b) Schematic of the forward network. The forward network has 4 fully connected hidden layers. The input matrix includes refractive index and wavelength-normalized geometric parameters of cylindrical meta-atoms, and the output is corresponding phase($\Delta\phi$) or transmission($T$) response. At another version, operation wavelength is also included in input matrix as an explicit parameter. Same architecture is used to train both phase and transmission models.

## Results and Discussion

**Cylindrical Meta-atoms.** The generic geometry of sub-wavelength cylinders on top of a transmissive substrate (Figure 1a) is chosen as the unit cell structure for investigation. Cylindrical dielectric meta-atoms can support both electric dipole (ED) and magnetic dipole (MD) resonances[33] enabling *0* to *2π* phase coverage and directional scattering. These meta-atoms can be approximated as waveguides or scatterers depending on their radius (*R*), height (*H*), and the *R/H* ratio. Such cylindrical scatterers have been employed as fundamental building blocks of beam deflectors[34-36], lenses[1-3,5,8-10] and resonators[36]. The design parameters are defined as refractive index, radius, and height of these meta-atoms, where the optical response is defined as transmission and introduced phase shift.

**Wavelength Normalization and Dimensional Mismatch.** The dielectric nano cylinders can be treated as dielectric nano-antennas. The geometric scalability of antennas states that the dimensions of a dielectric antenna operating at a wavelength $\lambda_0$ can be scaled by a scaling factor of $\lambda/\lambda_0$ to conduct the same operation at an arbitrary wavelength $\lambda$ provided that the material properties does not change. This implies that the parameters deciding the operation of an antenna are not the dimensions (*D*) themselves but their ratio with respect to operation wavelength $D/\lambda_0$. Based on this phenomenon, we normalized radius and height values to the wavelength of operation for every single data point in the dataset. Such normalization automatically solves dimensional mismatch problem by converting 1x3 (parameter set: *n, r, H*) to 1x501/251 (spectral response: T($\lambda$)/ $\varphi(\lambda)$) multivariate regression problem to 1x3 (*n, $r/\lambda$, $H/\lambda$*) to 1x1 (T($\lambda$) or $\varphi(\lambda)$) multivariable regression problem. In other words, the whole network becomes a multivariable function $F(n, r/\lambda, H/\lambda)$= T($\lambda$) or $\varphi(\lambda)$, where T($\lambda$) and $\varphi(\lambda)$ are scalars. Based on this approach, material dispersion can also be addressed directly by converting *n* to *n($\lambda$)*, as the model(s) generate independent solutions at every single frequency point.

Here, the knowledge on wavelength of operation ($\lambda$) is embedded inside the geometric parameters but not directly fed as a feature. To examine effect of spectral information as an explicit parameter, we construct 2 versions of each model, where $\lambda$ is also explicitly included in the input matrix of the 2nd version as seen in figure 1b.

**Forward Networks and Model Evaluation.** As shown in the Figure 1b, we use a conventional fully connected NN containing 4 hidden layers with 16, 100, 100, and 16 neurons, respectively. The parameter set is subjected to the proposed preprocessing step. Separate networks are trained for phase and transmission responses. Mean-squared-error (MSE) is used as overall performance metric. The transmission network achieved MSE of $7.2 \times 10^{-4}/7.3 \times 10^{-4}$ and the phase network achieved MSE of $2.1 \times 10^{-3}/2.1 \times 10^{-3}$ for training/test scores. Integrating operation wavelength to the input matrix improved MSE values to $7.1 \times 10^{-5}/8.1 \times 10^{-5}$ of $9.1 \times 10^{-4}/9.4 \times 10^{-4}$ for training/test scores of transmission and phase networks, respectively. The high and consistent training and test scores indicate that the regression problem is solved successfully without overfitting. Overall, the forward models are remarkably successful at predicting both the resonance and the off-resonance behavior.

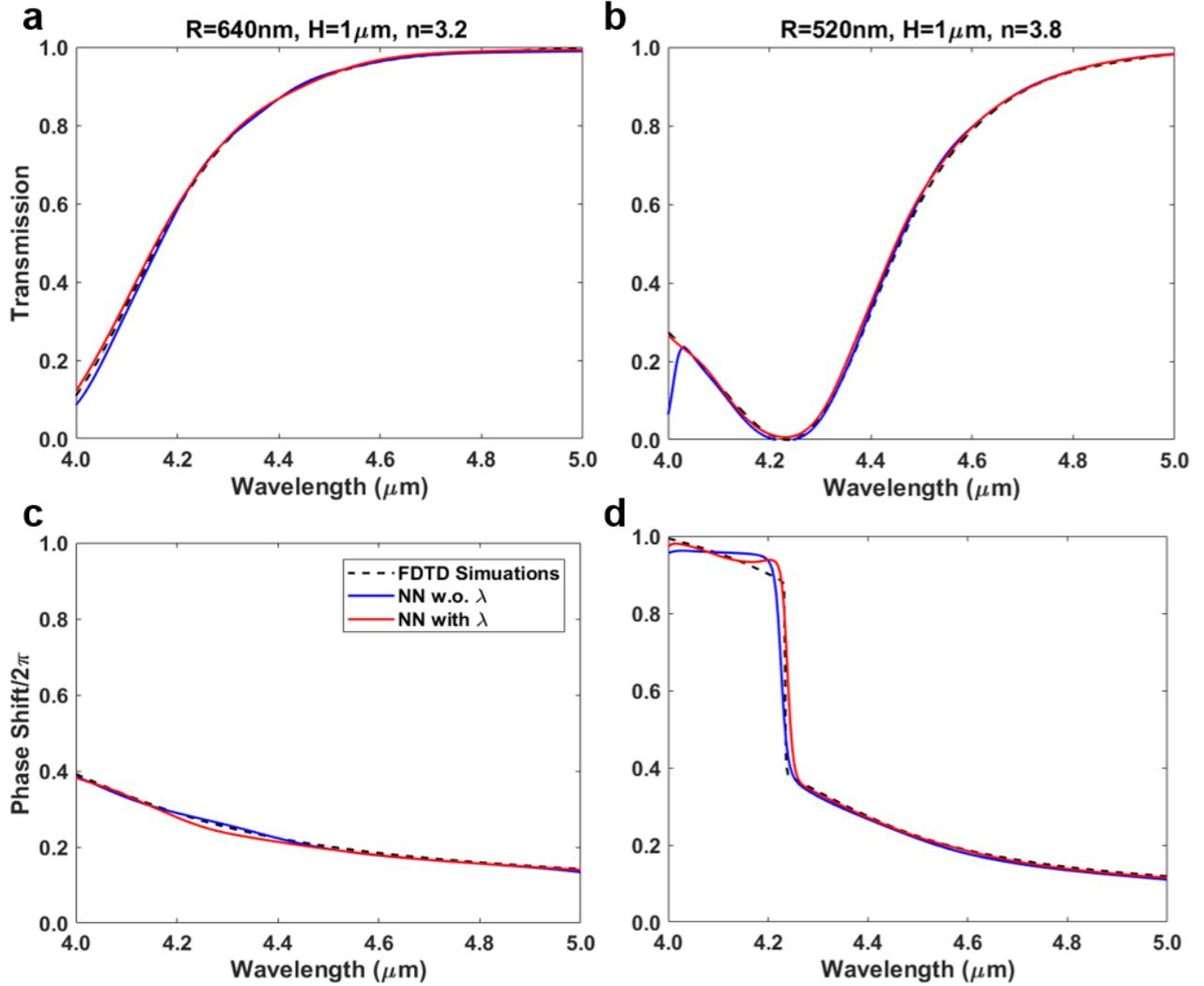

**Figure 2.** Comparison of predicted (by the models with and without λ) and simulated optical responses of exemplary samples from the validation set. First and second row indicates transmission and phase responses, respectively. The columns belong to two different parameter sets, where one illustrating off-resonance and the other illustrating resonance cases. Parameter sets are defined by their radius (*R*), height (*H*), and refractive indices (*n*).

Here, in order to fully represent the performance of our models, we selected 2 parameter sets from the unseen validation set. The selected sets exhibit off-resonance and on-resonance operation. Figure 2 compares the predictions of the both versions of NNs with FDTD simulations for these parameter sets. As seen in the Figure 2a,c, the predicted phase and transmission responses (red and blue lines) almost perfectly overlap with the simulated ones (black dashed line) at the off-resonance case. Figure 2b,d indicates performance of the NNs on-resonances, where transmission response exhibits an extremum point and phase response includes a sudden jump. As seen in the Figure 2b, both versions of models successfully resolve resonant dip. However, the 1st version, where $\lambda$ is not included in the input, the predicted response exhibits a sudden deviation from the simulation data near the left boundary of the wavelength range.

The reason behind this unexpected deviation can be found at the training data. Wavelength normalization of geometric parameters generates identical input sets (rows in the input matrix) that are corresponding different $\lambda$'s. As discussed earlier, geometric scalability of antennas suggests that the corresponding optical responses (rows in the output vector(s)) should also be identical. However, as we kept the unit cell side-length constant, the output data is not perfectly scalable. This imperfection is quite small, as indicated by MSE values at the order of $10^{-4}$ and seen in Figure 2, since the unit cell side length is large enough to ensure uncoupled operation along the entire parameter set. Nevertheless, one to many mapping problem arises, where identical input sets correspond slightly different outputs. Introducing $\lambda$ as an explicit input parameter solves this problem by differentiating these identical input sets. The resulted improvement can be observed at aforementioned improvement on MSE values and in the Figure 2b (See Supporting Information for more examples).

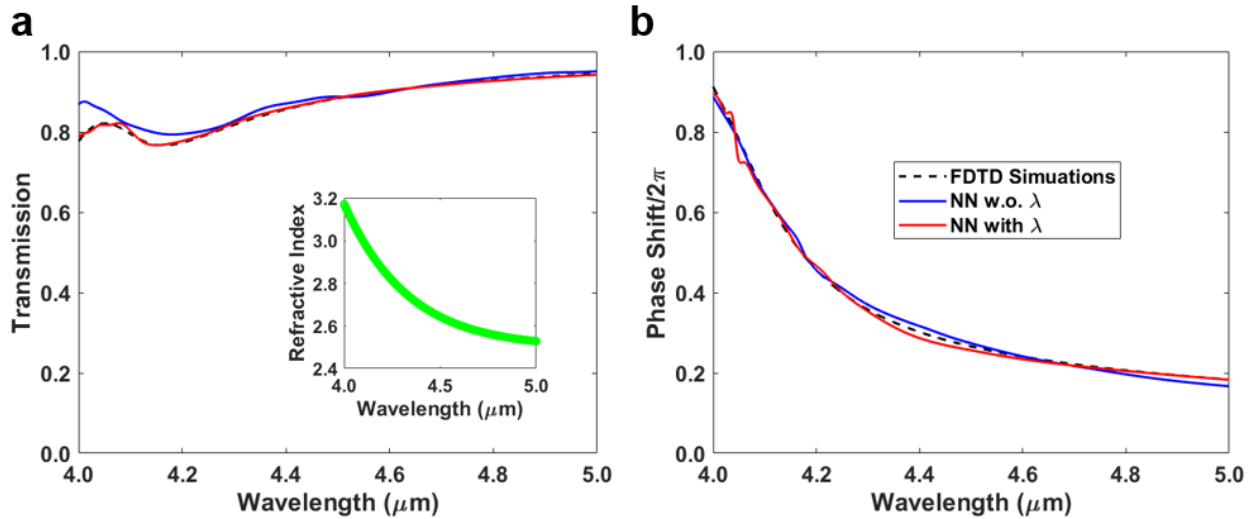

**Figure 3.** Comparison of predicted (by the models with and without $\lambda$) and simulated optical responses of a made-up dispersive material, where radius is 640 nm and height is 2 μm. a) Transmission, b) Phase response. The inset of a) shows the refractive index of the made-up material.

The proposed approach, as it divides broadband problem into single wavelength portions, offers an effective solution for dispersive materials, also. To investigate our models' performance in the case of dispersion, we made up a theoretical highly dispersive yet lossless material, where *n* exponentially decrease with wavelength from ≈3.2 to ≈2.5 as seen in the inset of Figure 3a. As seen in Figure 3, both versions successfully predict the optical response of the made-up material. However, aforementioned one to many mapping problem of the 1$^{st}$ version (without $\lambda$) show up as a relatively low prediction accuracy at the transmission response as seen in Figure 3a.

**Spectral Generalizability.** Spectral generalizability is another immediate result of our approach. Scalability property implies that the optical response of the parameter set ($n$, $r_0$, $H_0$) at wavelength $\lambda_0$ should be same with the optical response of parameter set ($n$, $r'$, $H'$) at wavelength $\lambda'$ if $r'/\lambda' = R_0/\lambda_0$ and $H'/\lambda' = H_0/\lambda_0$. As a result, since our networks are trained by these wavelength-normalized dimensions, their predictions should be spectrally generalizable. This property allows our models to make predictions outside their training spectral range. We note that our networks are trained considering uncoupled operation of cylindrical meta-atoms. To exploit spectral generalizability of our models, uncoupled operation should be ensured. A straightforward approximation to satisfy this condition is to consider wavelength normalized side lengths ($a/\lambda$) comparable with our training dataset, where $0.4 \leq a/\lambda \leq 0.5$.

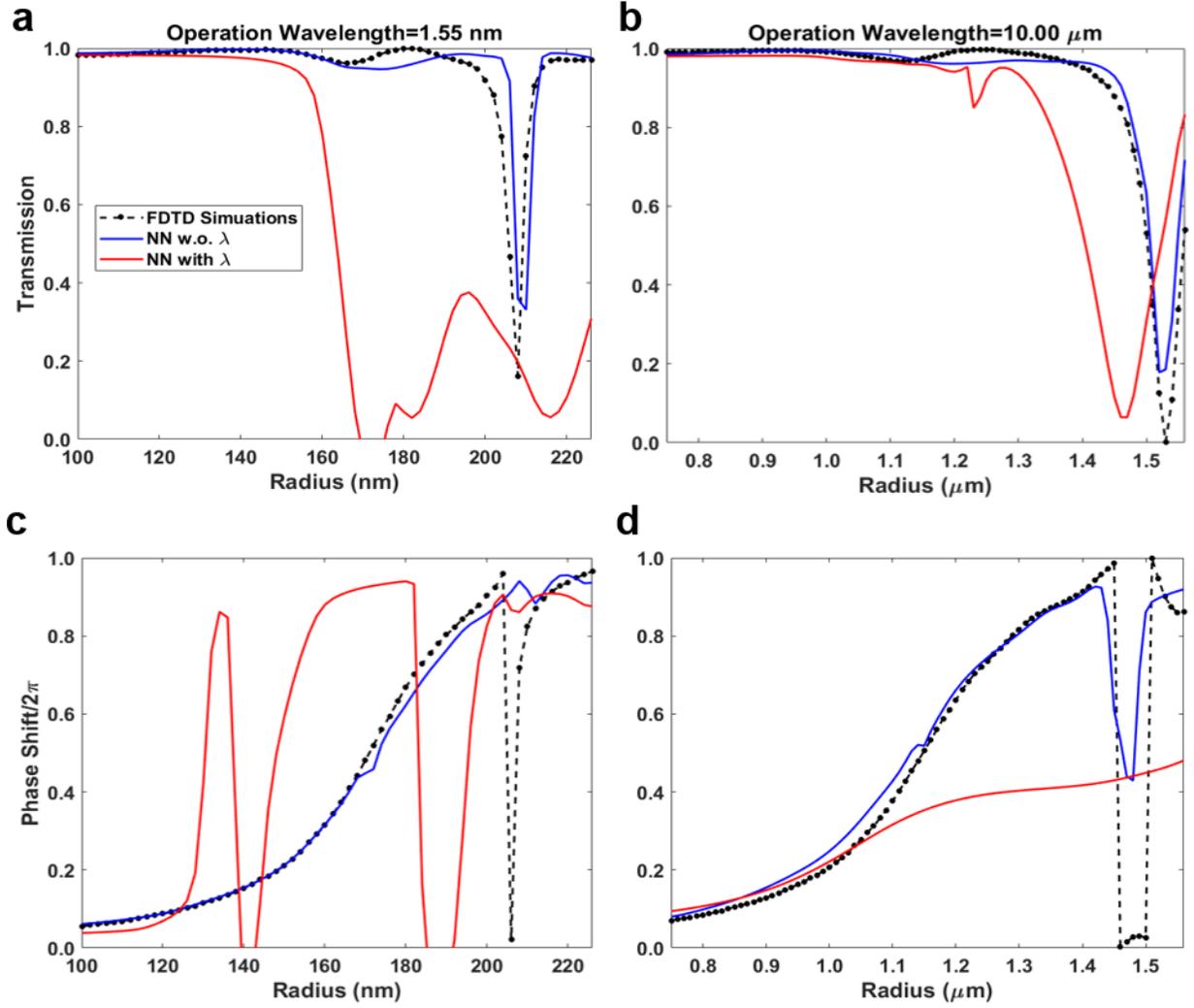

**Figure 4.** Comparison of predicted (by the models with and without λ) and simulated optical responses of test sets (1st (a, c) and 2nd (b, d)) each operating at a different spectral region (1.55 μm and 10 μm). First and second row indicates transmission and phase responses, respectively. Both sets employ silicon (Si) cylinders of fixed height and varying radius on top of a low index substrate ($SiO_2$ and $CaF_2$). The corresponding fixed height values are 0.94 μm and 5.50 μm and unit cell side lengths are 800 nm and 5 μm, respectively.

To investigate this property of our forward model(s), we use 2 parameter sets with operation wavelengths at near-IR, and far-IR. The first of these sets is taken from literature[4], and the last one is generated by us to be used as a test case for this study. These parameter sets are distinguished from the validation set used in Figure 2 in three ways; (i) they are operating outside the training spectral range of our model, (ii) real material models are employed in these test sets, and (iii) they present the optical responses of unit cell sets for single operating wavelengths.

Figure 4 shows the performance of our model(s) on these test sets. As seen in Figures 4a and 4b, the 1st version of our NNs (blue line) successfully predicts the transmission of the unit cell libraries. Although it cannot fully resolve the magnitude of the resonances, the model predicts the spectral position of resonances with a slight deviation outside the training spectral range. As it were for the validation set, the phase model (blue line) successfully predicts off-resonance behavior. However, it fails to resolve the resonances outside the training spectral range, as seen in the Figure 4c,d. The overall performance on both the transmission and phase prediction indicates that the NNs are already ready to be used as a pre-calculation tool even outside the training spectral range. They can save large amounts of time and computational power.

To complete our investigation on spectral generalizability property, we also test the 2nd version of our model(s) that are using wavelength as a parameter. This strategy had slightly improved the models' performance on the validation set. However, despite the in-range case (Figure 2), the model(s) that are directly employing wavelength as an input parameter (red line) drastically failed outside the training spectral range, as seen in Figure 4 (See Supporting Information for more details). This result indicates that the NNs did not learn the physical relation between optical response and wavelength of operation, and the spectral generalizability is completely based on the spectral scalability property.

**Inverse Network.** The inverse design is more complicated than the forward problem as the same optical response can be obtained from several different parameter sets. This problem must be overcome to solve the inverse problem. Here, we defined our inverse design task based on a common scenario, where the designer has given a certain material (or a limited number of options) and a target optical response. The design parameters to be determined are the radius and height of the cylindrical meta atoms. To solve the non-uniqueness problem, we construct an inverse network using modified version of previously shown tandem learning approach[19].

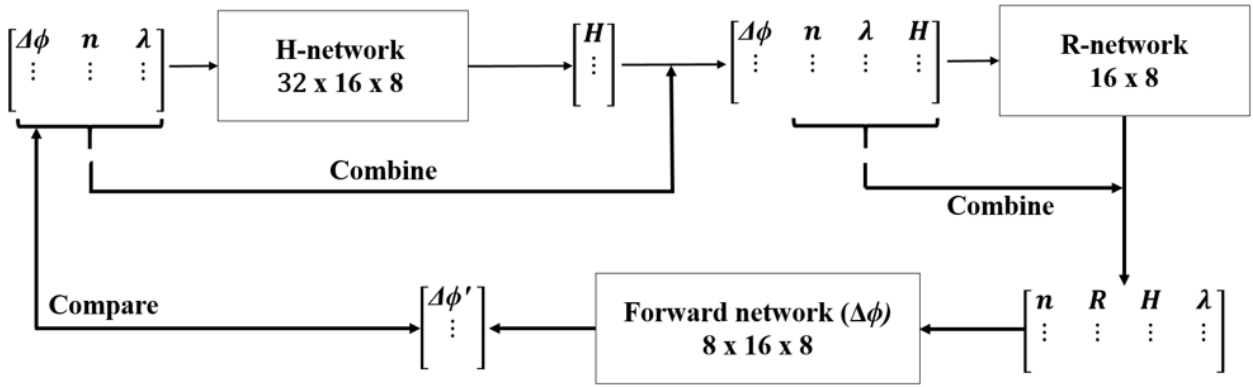

**Figure 5.** The complete inverse network structure. The R-network and the Forward network are pre-trained independently before training of thw whole network. During the training of the complete network, the wieghts of R- and Forward networka re fixed and the weights of H-network are altered to minimize the cost. The cost is defined as mean square difference between $\Delta\phi$ and $\Delta\phi'$.

We separated the two design parameters, height and radius. Assuming we know one of them, which is height in our case, we trained an inverse network to determine the corresponding radius values for given design specifications (n, λ, φ/T) and height. As the solution of this simplified problem is generally unique, if exists, in our limited training parameter space, uniqueness is not a problem for the R-model. Determining height values is a harder task and cannot be efficiently solved by the same method as there is no knowledge about the radius values. To solve this problem, we constructed the network structure shown in Figure 5. The first part(H-model) is a traditional inverse network, where the design specifications are taken as input and the output is the corresponding height. Combination of input and output of this network constitutes the input of the pre-trained R-model, and the corresponding radius is obtained. Then, we construct an input matrix for the pre-trained forward model by combining the output of R- and H- models with operation wavelength and the refractive index of material, which are given as design parameters. Finally, the weights of the H-network are trained to minimize the difference between the desired spectral response and the prediction of the forward network. Thus, the H-network becomes indirectly related to R-network. This full-network provides R-H sets for desired optical responses together with their predicted performance.

Using this network, we designed a unit cell library for a Ge (n≈4.015) metalens with an operation wavelength of 5 μm, as a proof of concept demonstration. The material and operation wavelength are intentionally chosen to be at the borders of training parameter space to fairly represent the capability of the inverse model. The desired phase response of each unit cell is defined as

$$\phi_{req}(r) = -\frac{2\pi}{\lambda}\left(\sqrt{r^2 + f^2} - f\right) \quad (1)$$

where *r* is the radial distance from the center of the lens, *λ* is the operation wavelength and *f* is the focal length. Diameter and focal length of the metalens are set to be 100 μm. Given these design specifications, the inverse network determines the parameter sets together with predicted optical response. The comparison of desired, predicted, and simulated phase responses and transmissions of unit cells are shown in Figure 6a. As seen in the figure, the simulated phase shift is closer than the predicted one to desired phase shift for small *r*, where the desired phase shift close to *2π*. This is a result of aforementioned under-prediction of the forward phase model around *2π*.

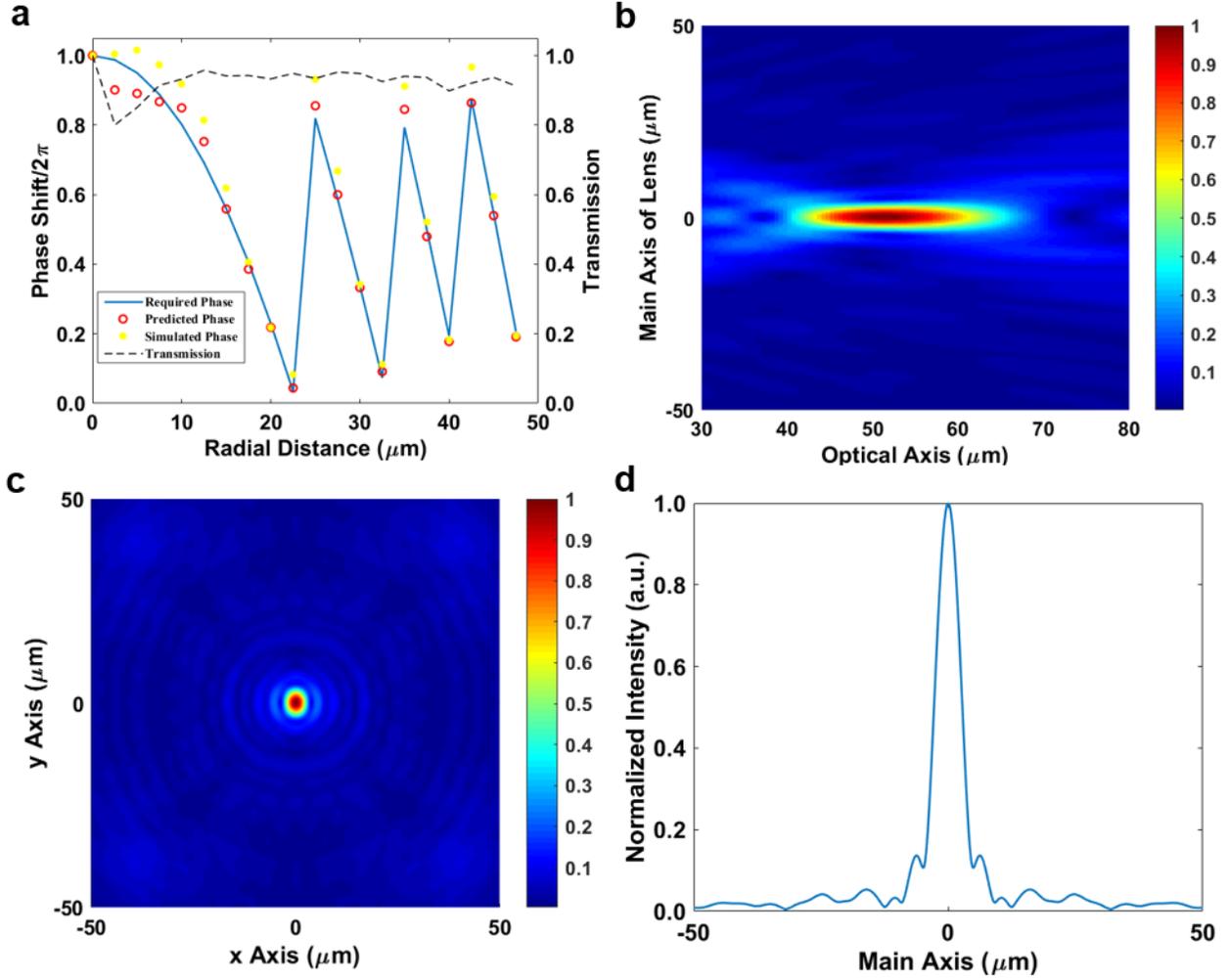

**Figure 6.** Design and performance of Ge metalens. a) The comparison of desired, predicted, and simulated phase responses together with simulated transmission response of the designed unit cells. b) Intensity profile in the focal region. c) Intensity profile in the focal plane at the focal distance (50 μm). d) Horizontal cut of the focal spot along the main axis. All intensity plots represent simulation results and the intensities are normalized to the peak value.

The overall focusing behavior of the resulted metalens can be seen from various perspectives in the Figure 6b-d. As seen in the Figure 6b, the metalens hits the target focal length. Figure 6c show the focal plane of the lens and the Figure 6d is its horizontal cut along the main axis. The full width half maximum (FWHM) is found to be $0.55\lambda$ (diffraction limit $\sim 0.50\lambda$), and the focusing efficiency is 70%. The overall performance of the metalens indicates that the proposed wavelength normalization approach can also be successfully applied to the inverse problem (see Supporting Information for further discussions on spectral generalizability and it is application with the inverse network).

**Conclusion**

In conclusion, we proposed a physics based pre-processing step solving dimensional mismatch problem and also providing spectral generalizability. We implemented the proposed approach to both forward and inverse problem for generic cylindrical meta-atom geometry, where we also address the non-uniqueness problem. As a proof-of-concept demonstration, we designed and simulated a Ge metalens with an operation wavelength of 5 μm. The proposed approach can directly be implemented to any geometry. Offering a general solution to dimensional mismatch problem and demonstrating spectral generalizability are critical steps towards realization of conventional learning based photonic models.

**Methods**

The dataset is generated by using full wave electromagnetic simulations based on FDTD method. Commercial simulation program Lumerical FDTD Solutions is used. Square unit cells with a fixed side length of 2 μm are used. The substrate thickness is taken to be infinite and the plane wave source is located inside the substrate modelling illumination from substrate side. The substrate is modeled as a low-index dielectric with refractive index of 1.45. The training spectral range is defined from 4 μm to 5 μm (Mid-wave infra-red (MWIR)), and resolved with 501/251 frequency points with equal wavelength spacing for transmission/phase response. The training parameter space defined in an effort to cover reported operation modes of dielectric nano cylinders (such as waveguide, high-contrast transmit array, resonant scatterer), and refractive indices of naturally available dielectrics (in this particular spectral range). The refractive index is changed from 2 to 4, the radius is varied from 300nm to 700nm, and the height is tuned from 750nm to 2500nm with step sizes of 0.2, 10nm and 250nm, respectively. This corresponds to a total number of 11x41x8=3608 samples. 1/8 of these samples are randomly selected and extracted as validation set, and the rest is used in training of the forward model.

For both the forward and inverse network, including the pre-trained forward and R-models, Keras framework is used. We train all the models using Adam optimizer[37], mean squared error (MSE) as the loss function and 'tanh' as activation function. The number of hidden layers for all the model is kept being the same at {16, 100, 100, 16}. We run the training for a total of 200 epochs with an updated learning rate at half point. After 100 epochs, the learning rate is updated from 0.002 to 0.001. The training process for a model takes about 20 minutes in a personal computer equipped with NVIDIA GeForce 1080 GPU.

**Supporting Information:**

Supporting Information Available: Learning curves of the models; Additional examples from the validation set; Additional examples on spectral generalizability of forward network; Inverse design of a metalens operating outside the training spectral range. This material is available free of charge via the Internet at http://pubs.acs.org

**Author Information:**

Corresponding Author: Koray Aydin
*E-mail: aydin@northwestern.edu

**Acknowledgments:**

K.A. acknowledges support from the Office of Naval Research Young Investigator Program (ONR-YIP) Award (N00014-17-1-2425). The program manager is Brian Bennett. This work is partially supported by the Air Force Office of Scientific Research under Award Number FA9550-17-1-0348.

# Supporting Information:

## A Physics Based Approach for Neural Networks Enabled Design of All-Dielectric Metasurfaces

Ibrahim Tanriover[1], Wisnu Hadibrata[1] and Koray Aydin[1]

[1]Department of Electrical and Computer Engineering, Northwestern University, Evanston, Illinois 60208, United States

a) Corresponding author: aydin@northwestern.edu

# SI-1 Training Forward Models

Below are the figures of both training and validation score over 200 epochs with batch size of 2048. All models are trained using Adam optimizer, mean squared error (MSE) loss function and 'tanh' activation function. The drop in the MSE at epoch 100 is for the updated learning rate from 0.002 to 0.001. By updating the learning rate, the loss value is improved. However, further decrease in the learning rate does not seem to improve the training process.

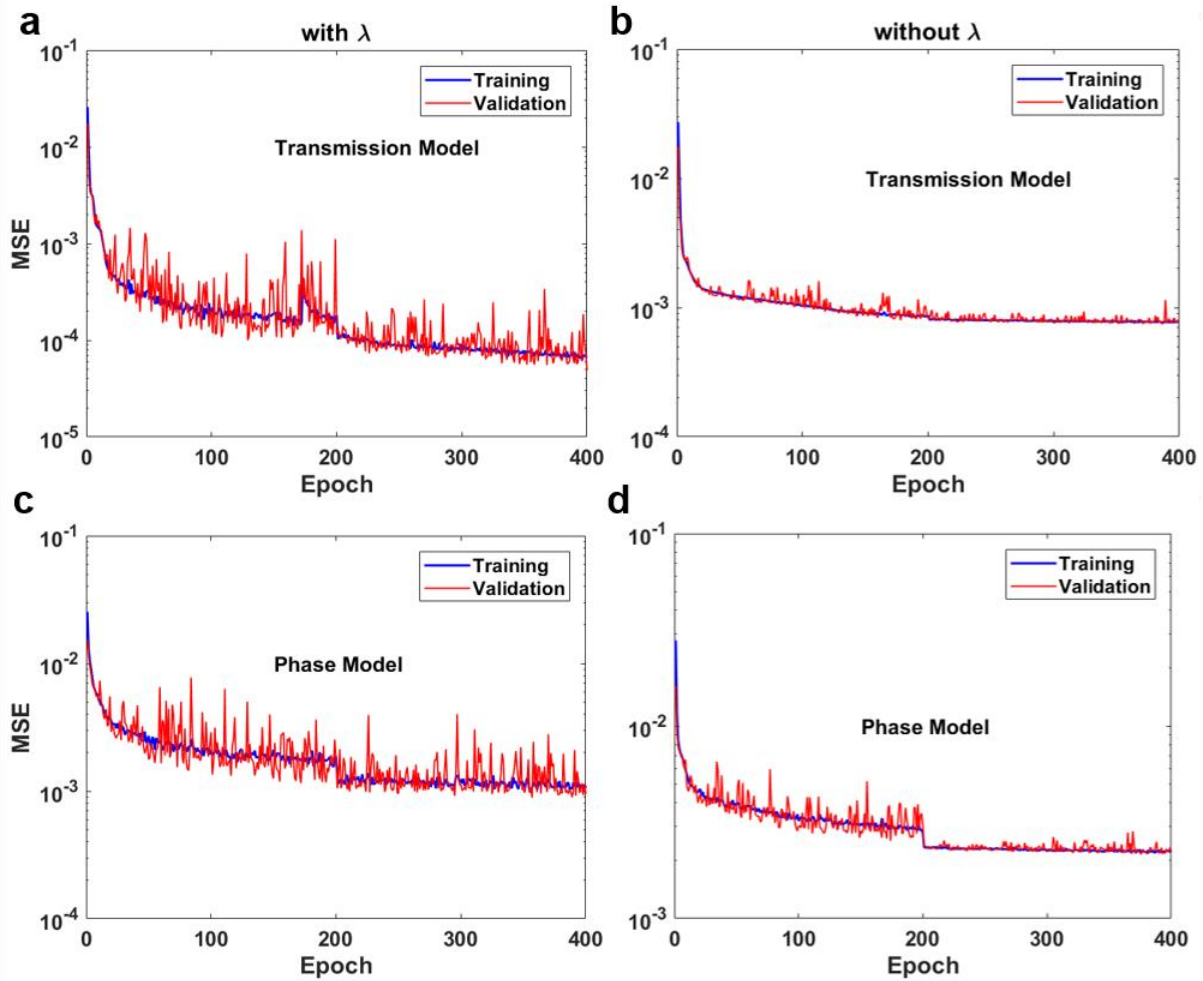

**Figure S1.** Learning curves of the forward models. a),c) Transmission and phase models employing operation wavelength ($\lambda$) as an explicit input parameter. b),d) Transmission and phase models where $\lambda$ is only embedded in wavelength-normalized geometric parameters

## SI-2 Additional Examples from the Validation Set

Supplementary figure S2 and S3 represent optical responses (transmission and phase shift, respectively) of several examples from the validation set. Each parameter set is defined by a radius ($R$), height ($H$) and refractive index ($n$) value.

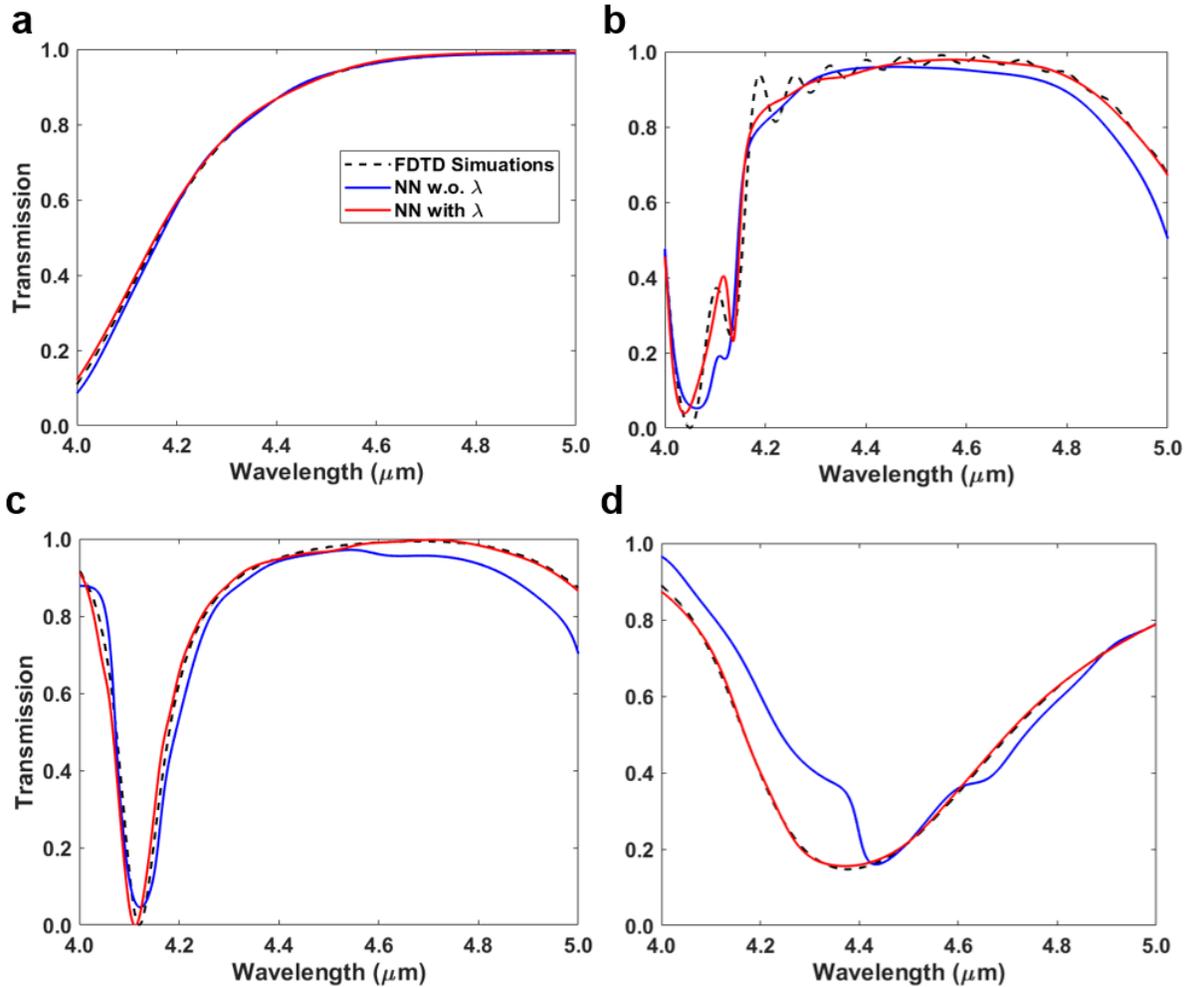

**Figure S2.** Comparison of predicted and simulated transmission values of several examples from the validation set. Corresponding radius ($R$), height ($H$) and refractive index ($n$) values are as follow; **a)** $R=640$ nm, $H=1000$ nm, $n=3.2$, **b)** $R=680$ nm, $H=1750$ nm, $n=3.8$, **c)** $R=660$ nm, $H=2000$ nm, $n=3.6$, **d)** $R=480$ nm, $H=1750$ nm, $n=3.8$.

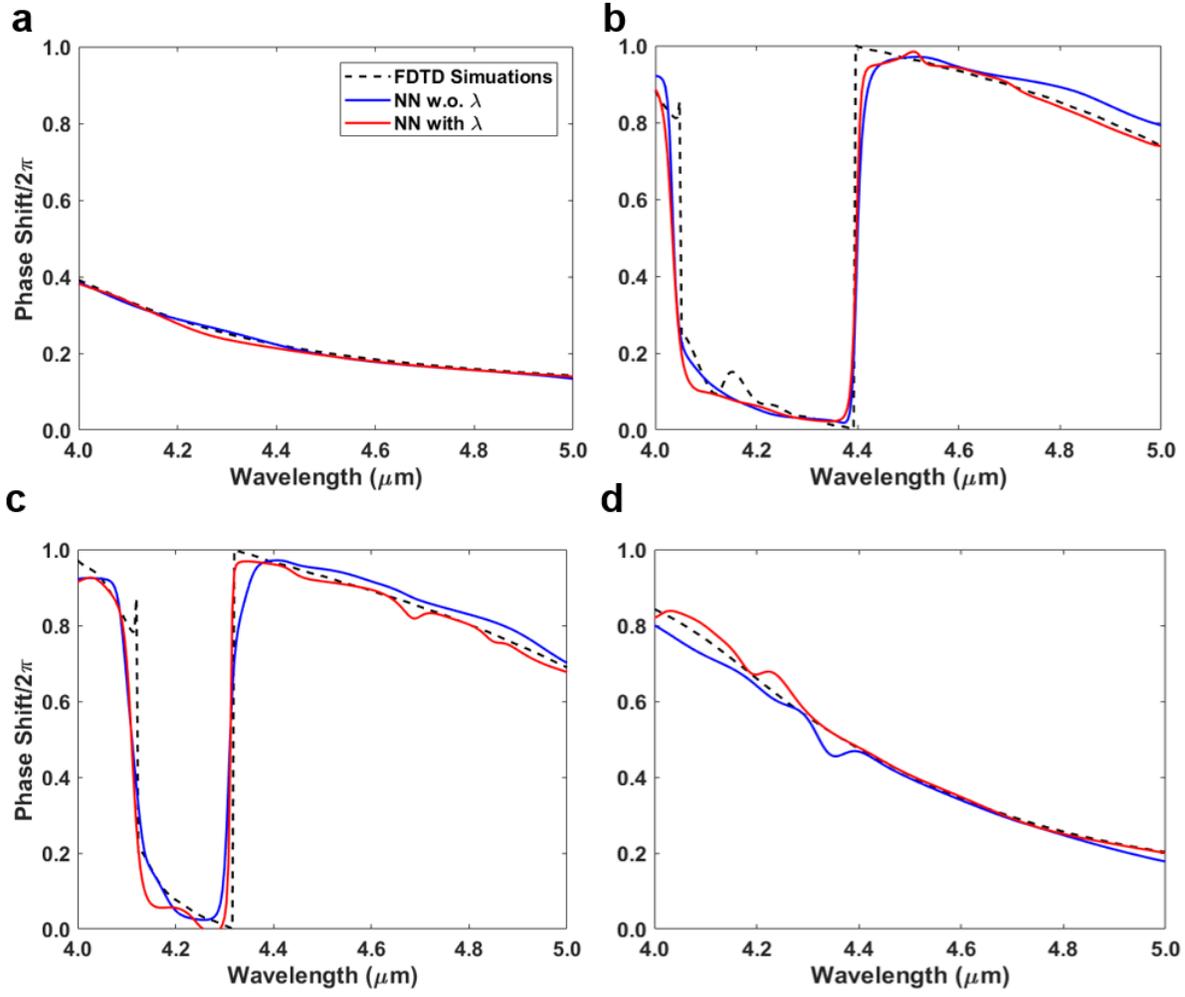

**Figure S3.** Comparison of predicted and simulated phase responses of several examples from the validation set. Corresponding radius (*R*), height (*H*) and refractive index (*n*) values are as follow; **a)** *R*=640 nm, *H*=1000 nm, *n*=3.2, **b)** *R*=680 nm, *H*=1750 nm, *n*=3.8, **c)** *R*=660 nm, *H*=2000 nm, *n*=3.6, **d)** *R*=480nm, *H*=1750 nm, *n*=3.8.

# SI-3 Additional Examples on Spectral Generalizability

## SI-3.a) Spectral Generalizability of Forward Network

Supplementary figure S4 and S5 represent optical responses (transmission and phase, respectively) of several unit cells that are operating between 9.0 and 11.0μm wavelengths. The unit cells are defined as Si nano cylinders on top of a low index ($CaF_2$) substrate, where cylinders have a height of 5.5μm and varying radius ($R$). Periodic boundary conditions are applied with a periodicity of 5.5μm. Here, the spectral generalizability is demonstrated for broad-band simulations despite the manuscript, where the spectral generalizability was demonstrated for single-wavelength simulation.

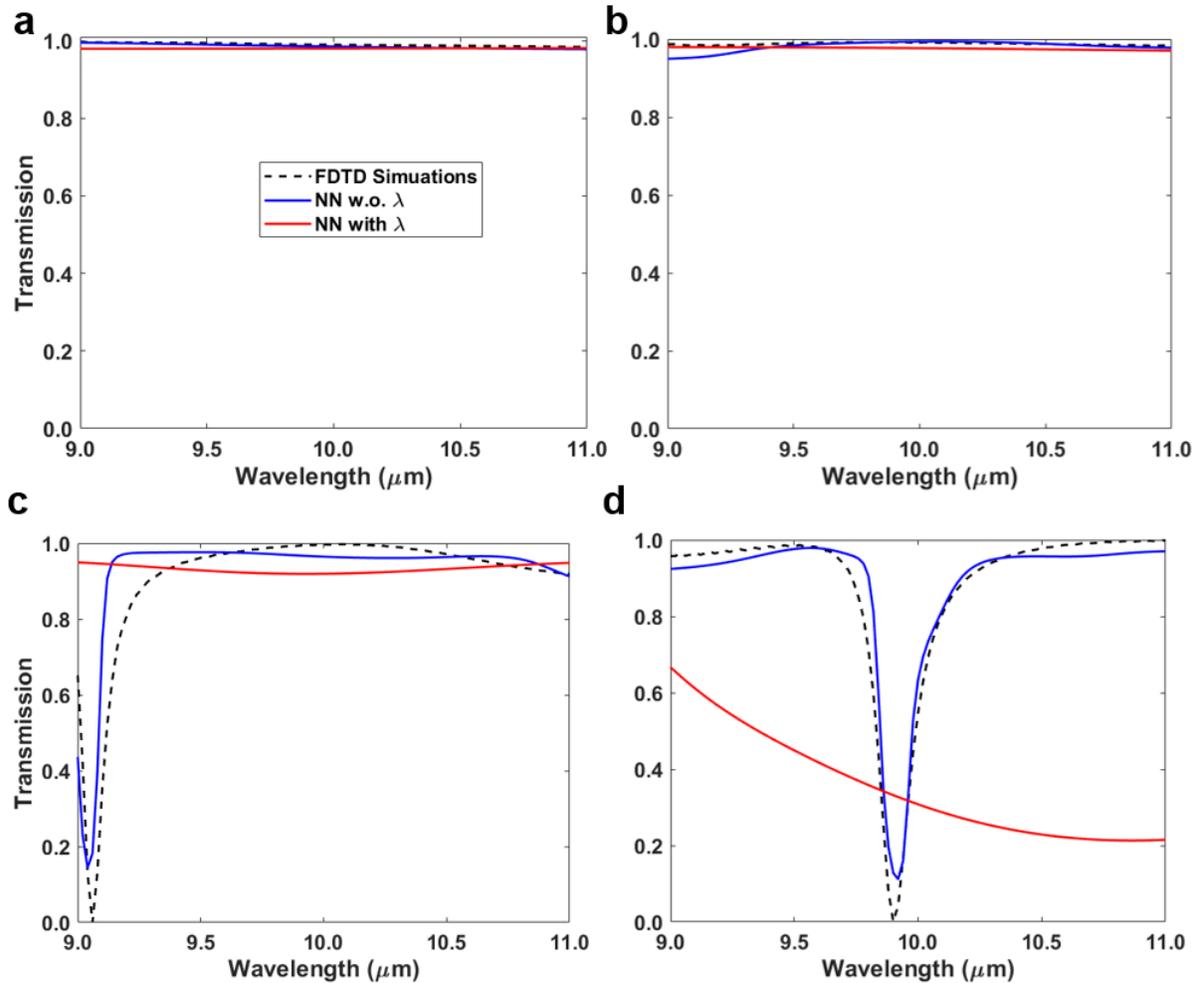

**Figure S4.** Comparison of predicted and simulated transmission values of several Si nano cylinders. Corresponding radius ($R$) values are: **a)** $R$=750 nm, **b)** $R$=1000 nm, **c)** $R$=1250 nm, **d)** $R$=1500 nm.

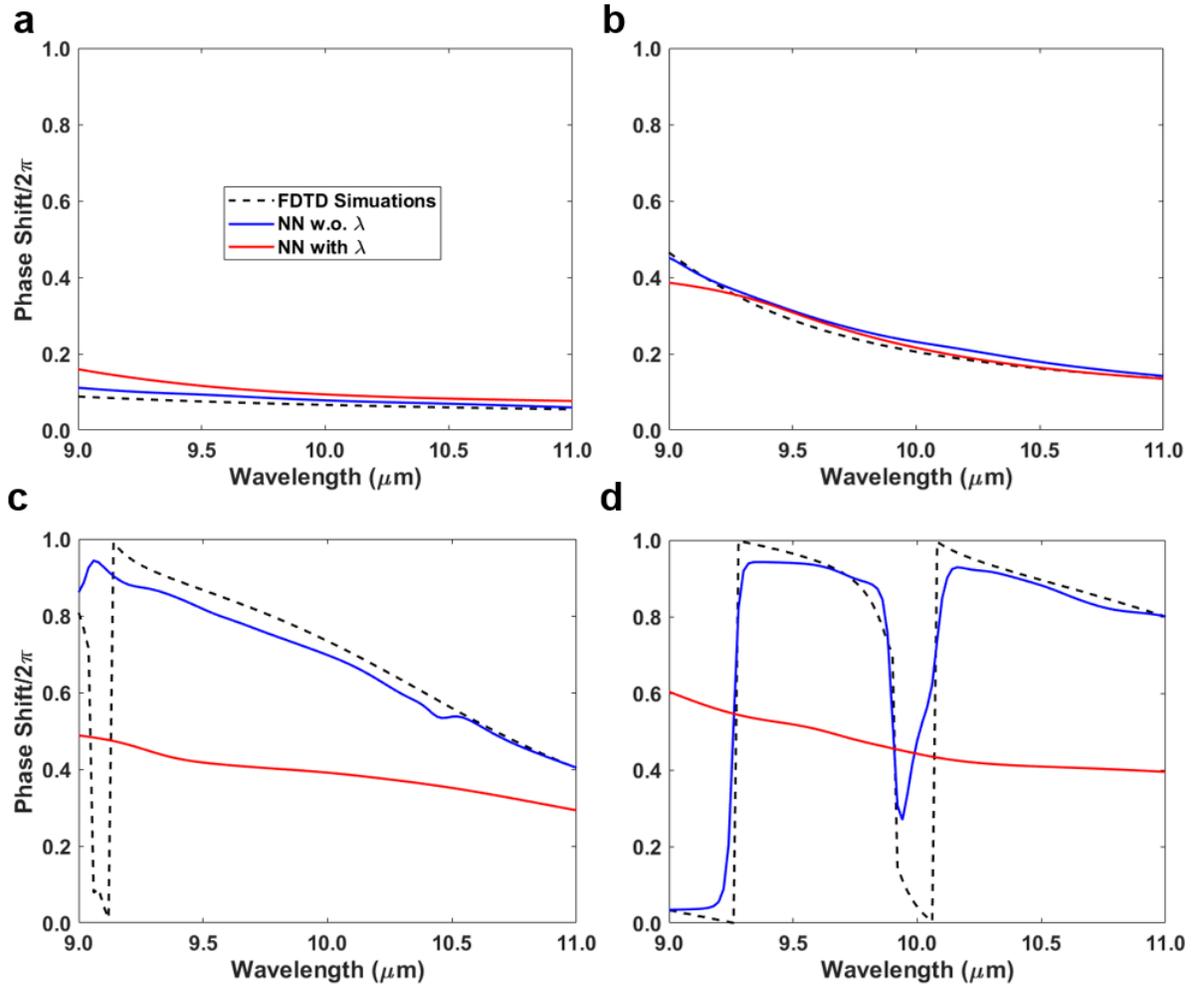

**Figure S5.** Comparison of predicted and simulated phase shifts of several Si nano cylinders. Corresponding radius (*R*) values are: **a)** *R*=750 nm, **b)** *R*=1000 nm, **c)** *R*=1250 nm, **d)** *R*=1500 nm.

Although it cannot resolve the full strength of the resonances, the transmission model without wavelength predicts the resonant points with a slight deviation (figure S4). The phase model without wavelength also performs successfully for the off-resonance case whereas it cannot resolve resonances as seen in the Figure S5. On the other hand, both the transmission and phase models with wavelength do not exhibit any sign of spectral generalizability, as discussed in the main script.

## SI-3.b) Spectral Generalizability of Inverse Network

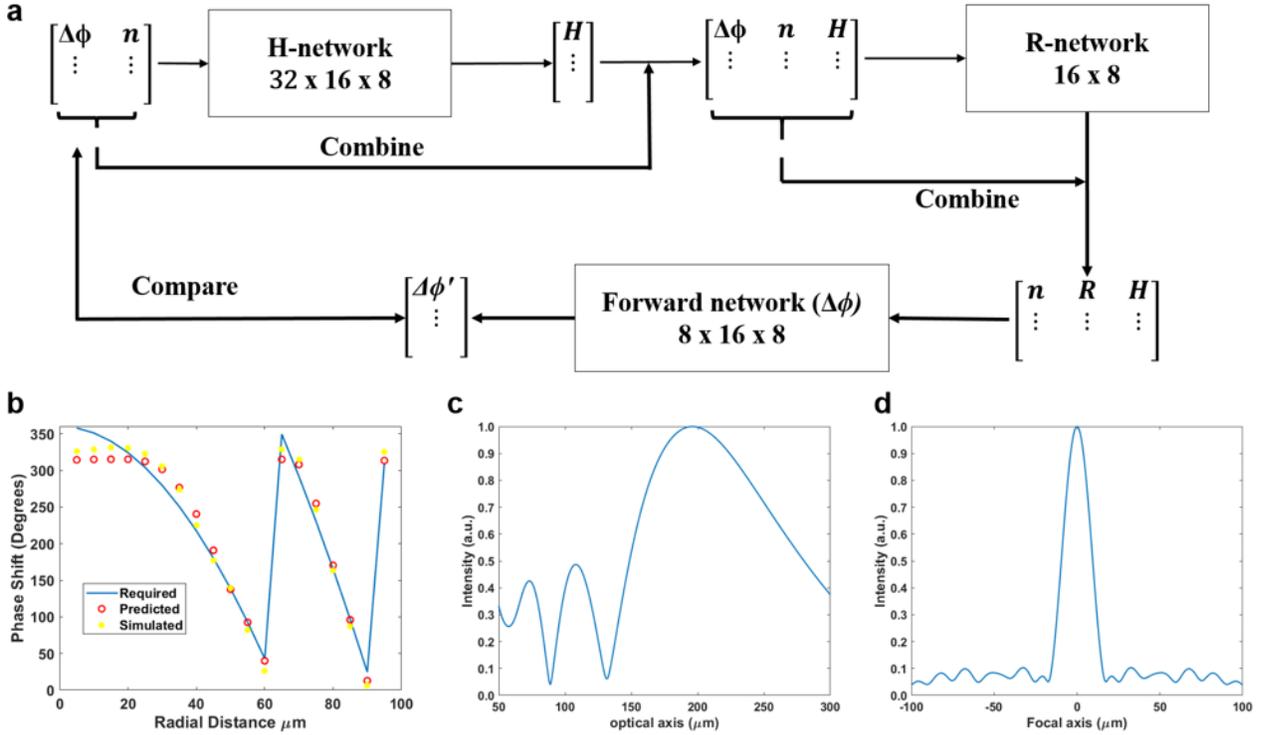

**Figure S6. a)** The spectrally generalizable inverse network **b)** Comparison of predicted, simulated and required (to build a metalens) phase profiles, where x axis is the radial distance from the center of the metalens **c)** Intensity of electric field after passing through the metalens along optical axis. The peak position corresponds to focal distance **d)** Intensity of electric field along the focal axis. All intensities are normalized.

Here, to further investigate spectral generalizability capacity of our models, we inverse-designed a metalens operating outside the training spectral range. Silicon (Si) was chosen as target material, 10 μm is chosen as operating wavelength, and the focal length is aimed to be 200 μm.

We employed a modified version of the inverse network that is discussed in the manuscript. As shown in figure S6(a), despite the inverse-network in the main text, this network doesn't take operation wavelength as an explicit input parameter, and uses the models without λ (wavelength) since these models are shown to have spectral generalizability.

The comparison of required and predicted phase profiles together with simulated ones are shown in figure S6(b). Figure S6(c) and(d) show intensity distribution along optical axis and focal axis, respectively. As seen in figure S6(c), the intensity peaks from 200 μm after the lens indicating that the aimed focal distance is achieved. Furthermore, intensity profile shown in figure S6(d) exhibits focusing behavior. Overall, figures S6(b), (c) and (d) indicates that the inverse network constructed based on our wavelength normalization approach operates successfully outside the training spectral range.